\documentclass[aps,prb,twocolumn,showpacs,floatfix]{revtex4}
\usepackage{graphicx}

\begin{document}

\title{Second-layer nucleation in coherent Stranski-Krastanov growth of
quantum dots}

\author{Jos\'e Emilio Prieto}
\email{joseemilio.prieto@uam.es}
\affiliation{Centro de Microan\'alisis de Materiales, \\
Dpto. de F\'\i{}sica de la Materia Condensada and Instituto 
Universitario ``Nicol\'as Cabrera'', 
Universidad Aut\'onoma de Madrid, E-28049 Madrid, Spain}

\author{Ivan Markov}
\email{imarkov@ipc.bas.bg}
\affiliation{Institute of Physical Chemistry, Bulgarian Academy of Sciences,
1113 Sofia, Bulgaria}

\date{\today}
\begin{abstract}
We have studied the monolayer-bilayer transformation in the case of the coherent
Stranski-Krastanov growth. We have found that the energy of formation of a
second layer nucleus is largest at the center of the first-layer island and
smallest on its corners. Thus nucleation is expected to take place at 
the corners (or the edges) rather than at the center of the islands as in
the case of homoepitaxy. The critical nuclei have one atom in addition to a
compact shape, which is either a square of $i\times i$ or a rectangle of
$i\times (i-1)$ atoms, with $i>1$ an integer. When the edge of the initial
monolayer island is much larger than the critical nucleus size, the latter
is always a rectangle plus an additional atom, adsorbed at the longer edge,
which gives rise to a new atomic row in order to transform the rectangle 
into the equilibrium square shape.
\end{abstract}


\maketitle

The Stranski-Krastanov (SK) mode of epitaxial growth is a nice example of an
instability of pla\-nar, two-dimensional (2D) growth against three-dimensional
(3D) islanding due to a non-zero lattice misfit between
the deposit and the substrate materials. This leads to the formation of 
arrays of self-as\-sembl\-ed small crys\-tal\-lites. In the case of
semiconductor overgrowth, these are known as quantum dots and have important
applications in optoelectronic devices. The physical
reason for the ocurrence of this 2D-3D transition is well established as the
gain of strain energy at the expense of surface 
energy.~\cite{Ter93,Ter94,Pol00,Dup98,Mul00,Bie01,Sch02,Joy02}
However, the mechanism of formation of 3D islands on the planar wetting 
layer in the case of coherent (dislocationless) SK growth is still an unsolved
problem in spite of intensive studies in the last two decades.

Voigtl\"ander and Zinner~\cite{Voi93} observed by scanning tun\-nel\-ing 
microscopy (STM) that faceted 3D Ge islands form at the same locations 
on a Si(111) surface at which 2D islands were observed 
in the initial stage of deposition immediately after exceeding 
the critical thickness of the wetting layer. Ebiko {\it et al.}~\cite{Ebi99}
found that the scaling function of the volume di\-stri\-bu\-tion of 3D 
InAs quantum dots on the surface of GaAs coincide with the scaling function 
for 2D submonolayer homoepitaxy with critical cluster size 
$i = 1$. Mo {\it et al.}\cite{Mo90} observed Ge islands representing
elongated pyramids (``hut" clusters) bounded by \{105\} facets inclined by
11.3$^{\circ}$ to the substrate. The authors suggested that the hut 
clusters are a step in the pathway to the formation of larger islands
with steeper side walls known in the literature as (rounded) ``domes" and
(faceted) ``barns".\cite{Sut04} Chen {\it et al.}\cite{Chen97} studied the
earliest stages of Ge islanding on Si(001) and established that Ge islands
smaller than the hut clusters are not bounded by discrete \{105\} facets. This
result was later confirmed by Vailionis {\it et al.}\cite{Vai00} who observed
the formation of 3 to 4 monolayers (ML)-high ``prepyramids" with rounded bases
which exist over a narrow interval of a Ge coverage in the beginning of the
2D-3D transition.  Also Arapkina and Yuryev found that the formation of the 
second layer of Ge clusters results in rearrangement of the first 
layer.~\cite{Ara10}
Sutter and Lagally\cite{Sut00} observed by low energy
electron microscopy (LEEM) the formation of an array of stepped mounds
(ripples) as precursors of the hut clusters on the surface of low
misfit alloyed SiGe films on Si(001) which are inherent to strained
films.~\cite{Asa72,Gri86,Sro89}

An insight concerning the formation of 3D islands on top of the wetting
layer came from Tersoff and LeGoues who suggested a nucleation
mechanism as the result of the interplay between the positive surface energy
of the islands and the relaxation of the strain energy in the islands relative
to that of the wetting layer.~\cite{Ter94} A critical volume is thus defined
beyond which irreversible 3D growth takes place. It was found that the
energetic barrier associated with the critical volume is proportional to
$f^{-4}$ where $f$ is the lattice misfit. This concept, although
very attractive, does not give any information about the mechanism of 
formation of the 3D islands. On the contrary, based on their 
observations Sutter and
Lagally suggested that 3D islands could be formed without the necessity of
overcoming an energetic barrier.~\cite{Sut00} Similar views and further
elaboration of the idea of barrierless trans\-formation of the ripples into
faceted islands were sug\-gested by Tromp {\it et al.}\cite{Tro00} 
and by Tersoff {\it et al.}\cite{Ter02}.

Priester and Lannoo, on the basis of microscopic calculations within the
Keating model, suggested that 2D islands appear as precursors of the 3D
islands,~\cite{Pri95}. Korutcheva {\it et al.}\cite{Kor00} and Prieto and
Markov\cite{Pri02} established on the base of 1+1 dimensional models that the
minimum energy pathway of the 2D-3D transition has to consist of a sequence
of states with thickness increasing by a single ML and that are stable
in separate intervals of volume. The first step in this process is the
rearrangement of monolayer into bilayer islands. This result was later
confirmed by a 2+1 dimensional model.~\cite{Pri05} Khor and Das 
Sarma~\cite{Kho00}
and Xiang {\it et al.}\cite{Xia10} reached the same conclusion by Monte Carlo
simulations. The first authors found that during growth, the material 
for the bilayer island originates almost completely from
the initial ML-high island in addition to some small amount 
coming from the vapor phase, the material for the 3~ML-island 
comes from the original bilayer island, etc. This layer-by-layer
mechanism of growth takes place if the deposited material is sufficiently
``stiff".\cite{Pri07} If the deposited material is soft enough such as 
Pb\cite{Gav99} or In,~\cite{Che08} the ML islands are expected to transform 
directly into multilayer islands with preferred heights (process known in the 
literature as ``electronic growth"\cite{Zha98}) because of the requirement 
of greater thickness to give rise to the necessary amount of strain 
relaxation, as was theoretically predicted in Ref.~[\onlinecite{Pri07}].

All these results are in accordance with the finding of Stoyanov and
Markov\cite{Sto82} who established for the case of Volmer-Weber growth at
zero misfit that an increase of the cluster volume gives rise to the 
stability of increasingly higher islands. The islands increase in height 
by one ML beyond certain critical volumes. 
The process of transformation of $n$-ML thick to ($n+1$)-ML thick 
islands is described by a curve giving the energy as a function of 
the number of atoms transferred to the upper level, which is 
characteristic for a nucleation process. It displays a maximum at a 
certain number of atoms and then decreases steadily up to the end of 
the transformation. The atoms of the ($n+1$)-th ML are detached from 
the edges of the lower monolayer island. The chemical potentials of 
the critical nucleus and the cluster underneath are equal, which is 
an indication of a true nucleation process. More details can be 
found in Ref.~[\onlinecite{Mar87}].

Recently, Villain noted that in highly-mismatched epi\-ta\-xy, second-layer 
nuclei are expected to form close to the edges of the initial 
ML-islands rather than at their centers.~\cite{Vil05} The reason is that 
the misfit strain is relaxed at the steps and 
``... atoms are happy to be there, because they find an atomic
distance which is closer to the atomic distance they would like to
have."~\cite{Vil05} Thus, for sufficiently large values of the
misfit, upper layer nuclei will form at the edges, while at small 
misfits, the adatom concentration will be highest at the island's center
and upper layer nuclei will form preferentially there, as in the case 
of homoepitaxial growth.~\cite{Mich04} The presence of Ehrlich-Schwoebel 
barriers for interlayer diffusion can also be important. This however leads  
to the formation of growth pyramids rather than to quasi-equilibrium 
3D crystallites.~\cite{Vil91}

\begin{figure}[ht]
\includegraphics*[width=4.2cm]{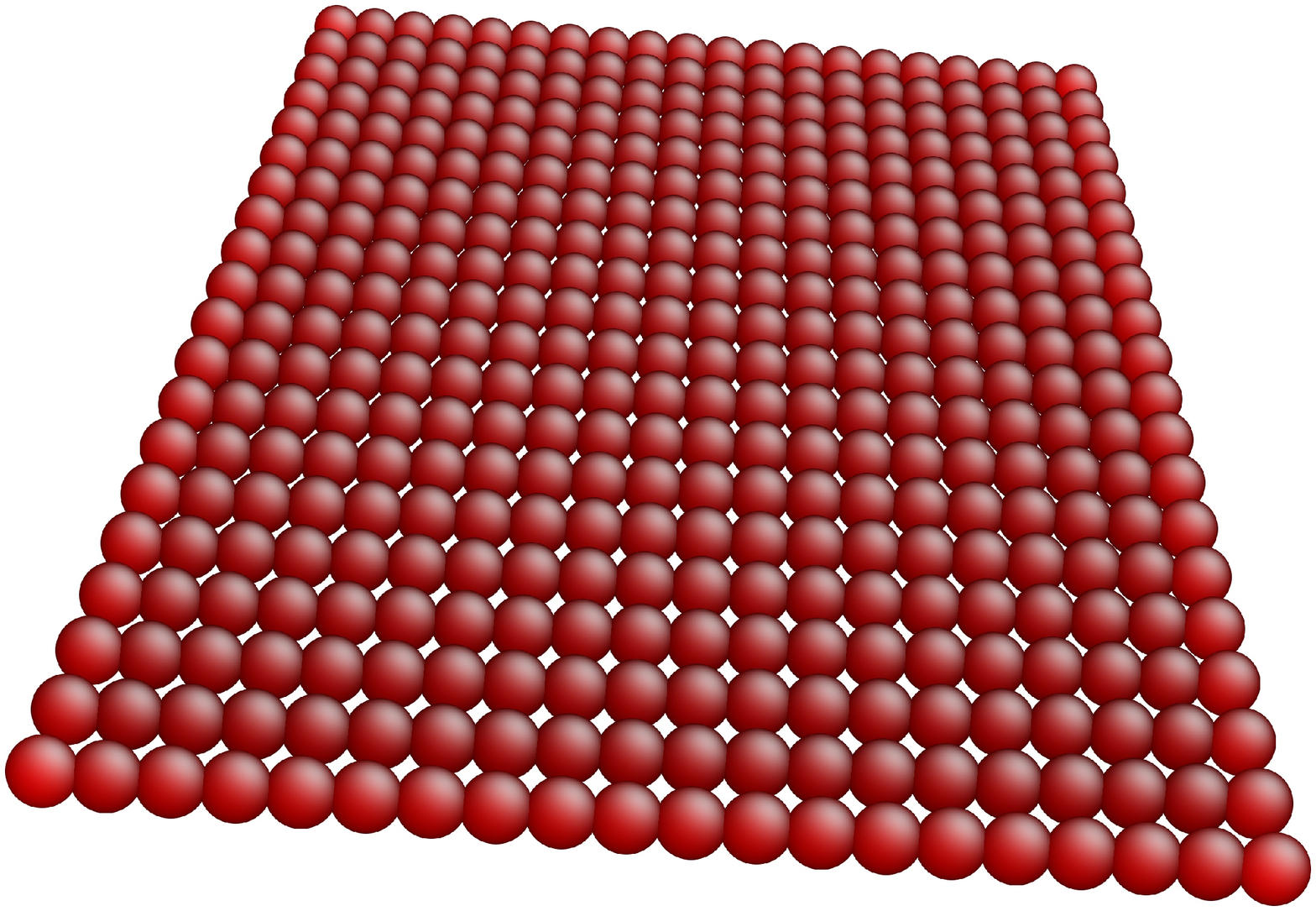}
\includegraphics*[width=4.2cm]{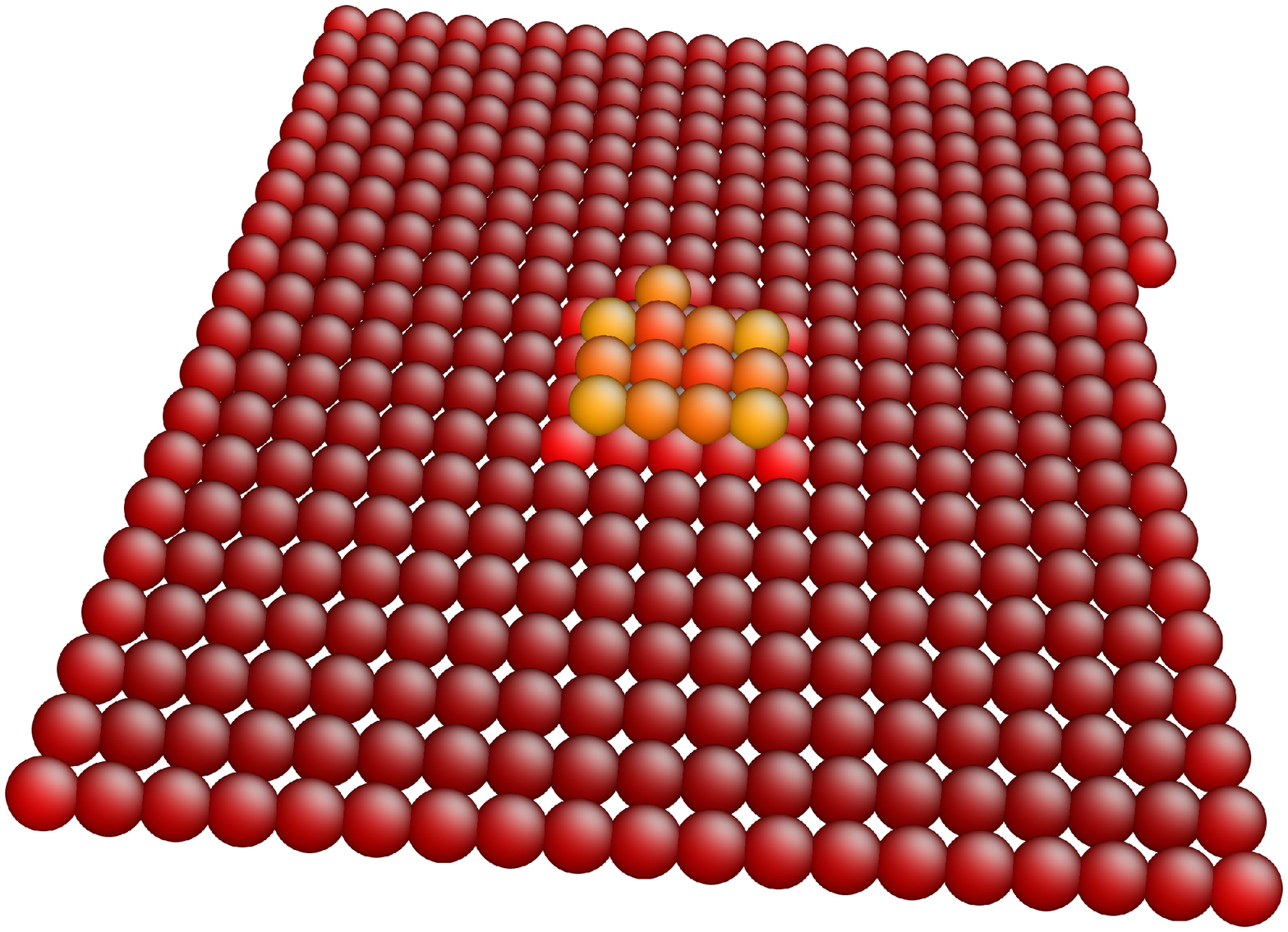}
\includegraphics*[width=4.2cm]{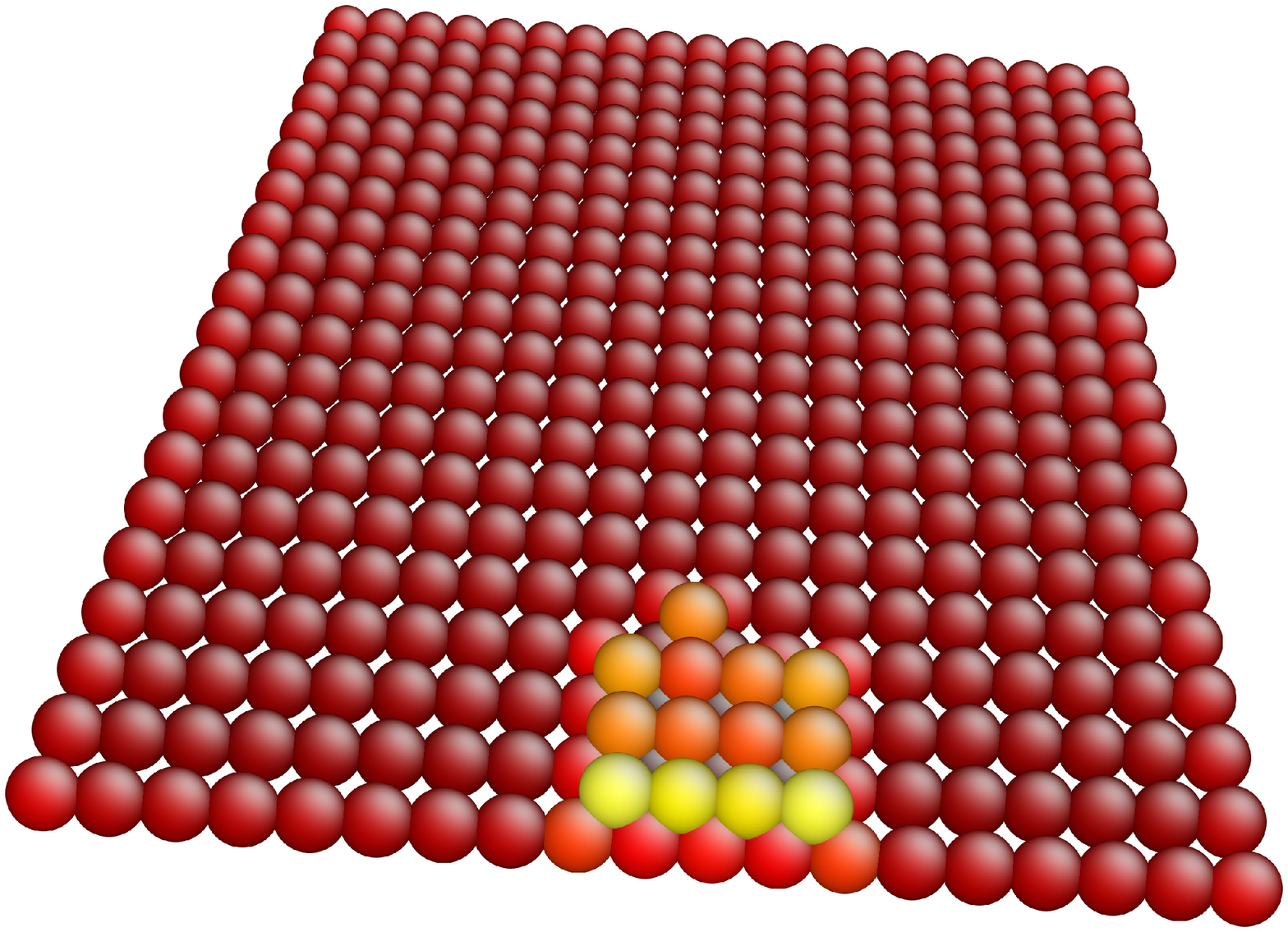}
\includegraphics*[width=4.2cm]{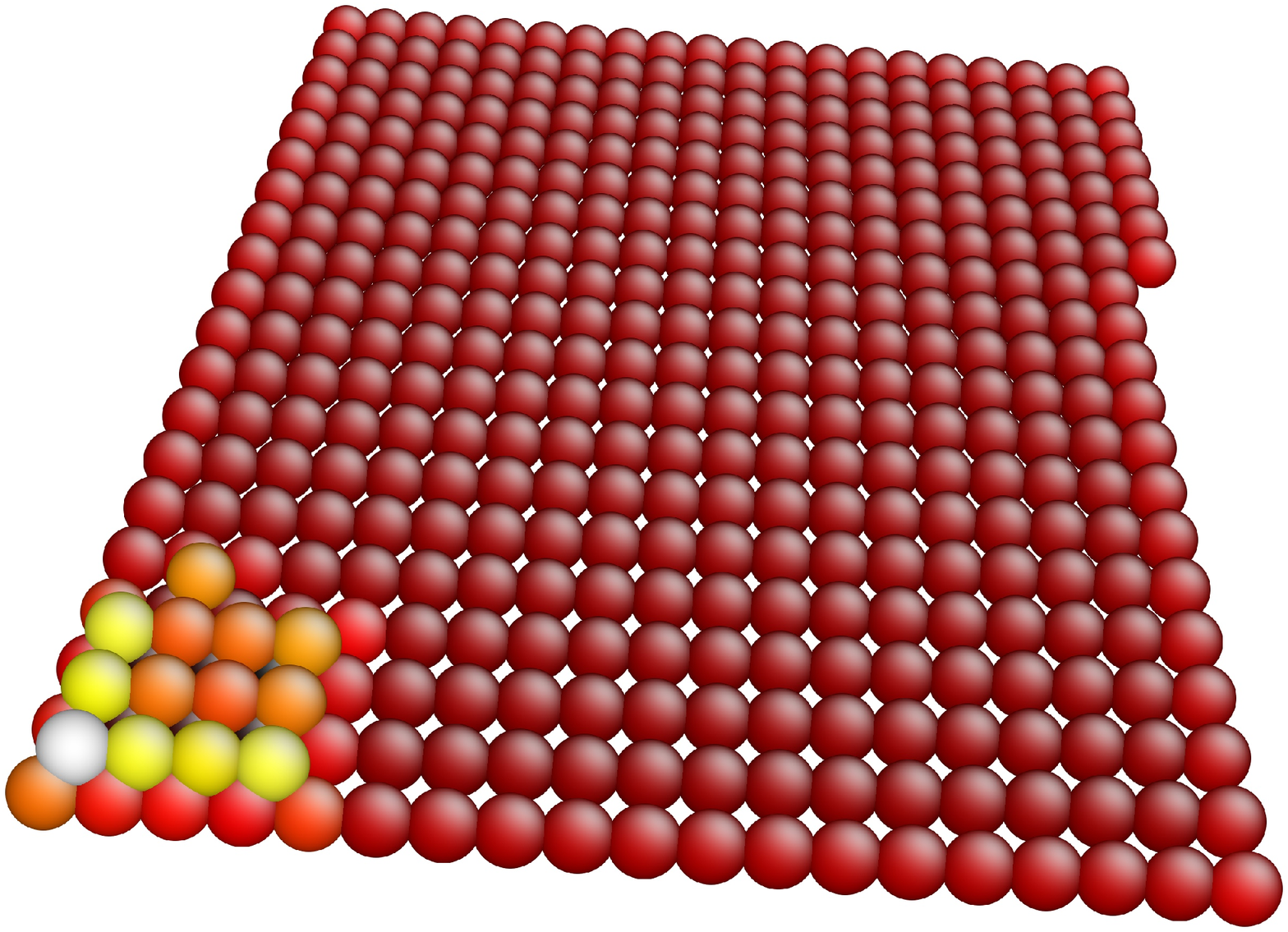}
\caption{\label{cecmodels} Locations of second layer nuclei. From top to
bottom and from left to right: initial 20$\times$20 monolayer island;
13-atoms second-layer cluster nucleated at the terrace center, at an
island edge and at an island corner of the initial monolayer island.
The color scale denotes the height of the considered atom and has
been represented using the ATOMEYE software.~\cite{Li} This height is
measured above the level of the corresponding crystallographic plane, but
a constant fraction of the interlayer distance has been added in order to
better distinguish atoms from different levels. The height is biggest at
edges and corners due to the atoms ``climbing up" on their neighbours
underneath due to strain relaxation. The lattice misfit is -7\%.}
\end{figure}

In the present paper we study the formation of second layer nuclei at
different locations on the first-layer island: center, edge
and corner (see Fig.~\ref{cecmodels}).
The model has been described in detail elsewhere.~\cite{Pri05} 
Briefly, we consider an atomistic model in $2+1$ dimensions, in which the 
3D crystallites have fcc structure and (100) surface orientation, thus 
possessing the shape of truncated squ\-are py\-ra\-mids.
The lattice misfit is the same in both orthogonal in-plane di\-rec\-tions. 
We consider interactions only in the first coordination sphere. Inclusion 
of further coordination spheres is not expected to alter qualitatively 
the numerical results as long as epitaxial structures remain coherent. We 
perform a simple minimization procedure. The atoms interact 
through a pair potential containing two adjustable pa\-ra\-me\-ters 
$\mu $ and $\nu $ ($\mu > \nu $),~\cite{Mar88,Mar93}
\begin{equation}
V(r) = V_0\Bigl[\frac{\nu }{\mu - \nu}e^{-\mu (r-b)} - \frac{\mu}{\mu - \nu}
e^{-\nu(r-b)}\Bigr],
\end{equation}
where $b$ is the equilibrium atom separation, For $\mu = 2\nu$ the above
potential adopts the familiar Morse form. In spite of its simplicity, the
above potential includes all necessary features to describe real materials
(bonding strenght and anharmonicity). The interatomic spacing of the 
substrate is $a$ so the lattice misfit is given by $f = (b-a)/a$. 
The substrate is assumed to be rigid.

We study the transformation of mono- into bilayer islands, considered as 
the first step of the 2D-3D trans\-formation, along the procedure 
de\-velop\-ed in Ref.~[\onlinecite{Sto82}].  
We assume the following model processes: The initial state is a square,
ML-high island, whose size is larger than the critical value and which 
is thus unstable against mono- to bilayer is\-land transformation.~\cite{Pri05}
Atoms detach from the edges of the initial ML-island, diffuse on top 
of it and eventually aggregate and give rise to second-layer nuclei.
These grow at the expense of the atoms detached from the edges of the lower
island up to the moment when the upper island completely covers the 
lower level. To simulate this process, we detach atoms one by one 
from the edges of the lower island and arrange them on top, forming 2D 
clusters as compact as possible at one of three different locations: 
At the center of the terrace (center), at the 
center of one edge (edge), and at one of the corners (corner) of the
initial monolayer island (see Fig.~\ref{cecmodels}). 
The second-layer clusters are always as compact as possible, i.e. either 
a square of $i\times i$ or a rectangle of $i\times (i-1)$ atoms 
(where $i>1$ is an integer number) plus eventually a fraction of an 
atomic row placed at a free edge of the island (at the longer edge in the
case of a rectangular island). This mechanism is expected to describe
most closely the experimental 
results,~\cite{Voi93,Ebi99,Mo90,Sut04,Chen97,Vai00,Ara10} as discussed above.

The change in energy associated with the process of transformation at a
particular stage is given by the difference between the total energies of 
the incomplete bilayer island and the initial monolayer island. We then
compare the nucleation barriers and the number of atoms in the nuclei formed
on the lower island center, edge and corner locations. We find that,
in addition to Villain's kinetic reasons, thermodynamics also plays in
favor of a preferred formation of second layer nuclei at the island's 
edges and corners where the strain is smaller.

As shown in Ref.~[\onlinecite{Pri05}], the sign of the misfit plays a 
cru\-c\-ial role in the mechanism of trans\-for\-mation of mono- to
bi\-lay\-er is\-lands. When
the second-layer clusters are formed at the center of the first-layer 
islands, the curves for positive misfits show a true nucleation 
mechanism. The curves display sharp maxima followed by a decrease of 
the energy up to the complete transformation. The value at the maximum 
gives the energy barrier for nucleation of the upper layer. 
In the case of negative misfit, we observe a totally different mechanism. 
The transformation energy does not display a well-defined maximum but 
increases steadily up to a (relatively high) value beyond which it 
steeply collapses, close to the end of the transformation. 
The collapse is due to the coalescence of the single steps into bilayer 
facets which possess a much smaller energy.

\begin{figure}[ht]
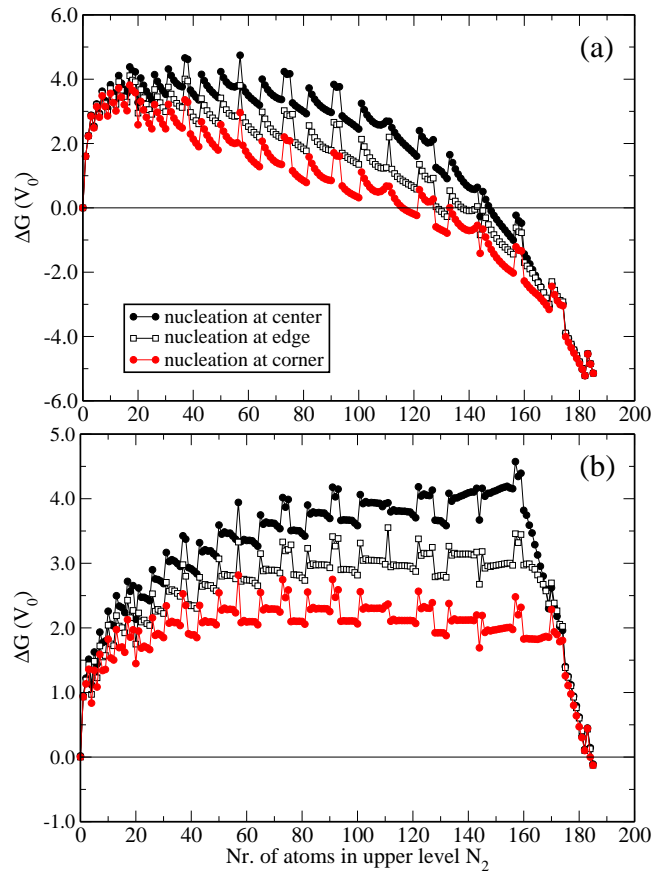

\includegraphics*[width=8.5cm]{./Fig2a.eps}
\includegraphics*[width=8.5cm]{./Fig2b.eps}
\caption{\label{trcurves} Transformation curves (total energy as a function
of the number $N_2$ of atoms transferred to the second ML) for (a)
positive (+3.5\%) and (b) and negative (-12.0\%) values of the misfit.
A potential with $\mu = 16$ and $\nu = 14$ was used in the simulations and
an initial monolayer island containing 20$\times$20~=~400 atoms was
considered. The second-layer islands are formed at the center, at one
of the edges and at one corner of the initial, monolayer island.}
\end{figure}

This behaviour, different for different signs of the lattice misfit, 
is confirmed in Fig.~\ref{trcurves} by the energy curves corresponding 
to nucleation at different locations on the initial square-shaped monolayer 
island containing 400 atoms. Figure~\ref{trcurves} shows transformation 
curves for second-layer islands nucleated at the center of the monolayer
island, at the middle of one of its edges and at one of its corners; 
the patterns of transformation assumed in the simulations
is comparable in all cases, in particular 2nd-layer clusters of similar 
shapes were chosen irrespective of their location on the initial 
monolayer island. 

Figure~\ref{trcurves} shows that, for both signs of the lattice misfit, 
the process of island nucleation at the corner position has the 
lowest energy barrier and the terrace-center position has the highest 
one, while the edge position shows an intermediate value. This 
can be understood in terms of the different level of strain 
relaxation at different positions of coherent islands. Atoms at the 
center of a terrace are forced by their lateral neighbours to adopt a 
similar lateral distance as atoms in the lower level and hence possess 
the highest strain.~\cite{Kor00,Pri05} In contrast, atoms at edges and 
even more 
at corner positions have the possibility to displace laterally due to 
their reduced lateral coordination, so they relax epitaxial strain 
more efficiently. In this way, the process of 3D clustering is favoured
by a higher degree of strain energy reduction in the cases of corner 
and edge nucleation. 
For negative values of the misfit (expanded overgrowths), the formation of 
second-layer islands on the first-layer islands corners shows a slightly 
different behavior compared with the growth of islands at the terrace 
center [Fig.~\ref{trcurves}(b)]. In addition to the expected collapse of the 
energy at the end of the transformation, the energy displays broad 
maxima at a relatively large number of atoms. Thus we observe a 
superposition of the nucleation-like behavior and the collapse of the 
energy due to the coalescence of the single steps. Increasing the 
absolute value of the misfit leads to a decrease 
of the number of atoms in the critical nucleus.

\begin{figure}[ht]
\includegraphics*[width=8.5cm]{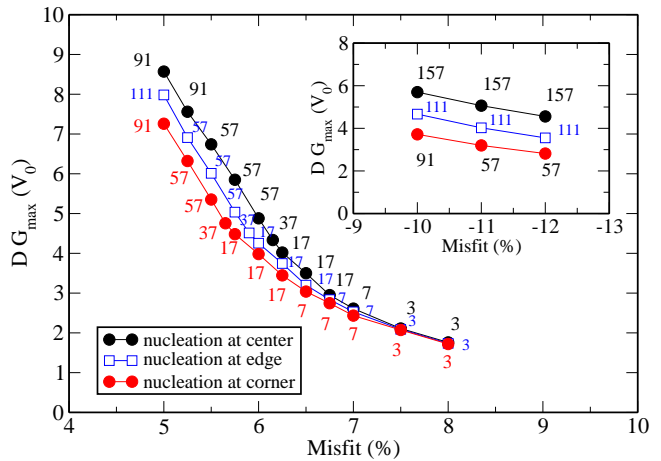}
\caption{\label{DGcurves} Heights of the nucleation barriers in units of $V_0$
as a function of the value of the lattice misfit (main plot: positive misfits;
insert: negative misfits). The figures at
each point denote the number of atoms, $n^*$, in the critical nucleus. The
values for compressed islands were calculated for $\mu = 2\nu = 12$,
those for expanded islands for $\mu = 16$ and $\nu = 14$. A cluster size of
400 atoms was considered.}
\end{figure}

Figure~\ref{DGcurves} shows our main result. It represents the heights 
of the nucleation barriers as a function of the lattice misfit both for 
positive and negative values and for the cases of second-layer nucleation at 
the center, edge and corner of the initial monolayer island. The figures 
close to the data points give the number of atoms in the critical nuclei. 

As seen in Fig.~\ref{DGcurves}, for the case of compressed islands, 
small critical nucleus sizes and
correspondingly small barrier heights are obtained for relatively small
values of the lattice misfit, no larger than 8\%. Both magnitudes increase
markedly with decreasing misfit value. In contrast, for expanded
layers, much larger, unrealistically high absolute values of the 
lattice misfit, between 10 and 12\%, are required to obtain barriers 
of comparable heights. The effect is even stronger when considering 
that the calculations for negative misfits were performed for a stronger 
potential. Furthermore, the
numbers of atoms in the critical nuclei is much larger than in compressed
islands and the dependence of critical sizes and barrier heights
on the absolute value of the misfit is also much less pronounced. 
The curves for nucleation at different locations show a clear energetic 
preference for corner as compared to center nucleation. As an example, 
for reasonably high 
positive values of the misfit (below 6\%), the difference in the 
barrier heights between center and corner second-level nucleation 
can be larger than the energy of a single atomic bond 
(V$_0$ in our model), representing a decrease of roughly 25\% or even more.
This is of enormous significance given the exponential dependence of the 
nucleation rate on the barrier height.
We conclude that for both signs of the lattice misfit and due to
thermodynamic reasons, nucleation is expected to take place preferentially 
at the islands corners, followed by the edges, rather than at the 
islands centers. 

Experimental evidence of the effect of misfit strain on second-layer 
nucleation has been obtained in organic [tetracene on H-passivated 
Si(001)\cite{Shi03}] and in metal-on-metal system such as 
Pd/Cu(001).~\cite{Mur95} A further convincing example is the case 
of SK growth of InAs$_x$P$_{1-x}$ nanowires on InP.~\cite{Mol07} 
Images show that the second-layer clusters form preferentally 
on the upper side of steps, where maximum strain relaxation occurs. 
According to the authors, the explanation of this observation 
does not require accounting for the Ehrlich-Schwoebel effect. 
An explanation for this behaviour can be of kinetic origin, based either 
on Villain's arguments or on the presence of Ehrlich-Schwoebel barriers 
(the adatom concentration has a maximum value at the step edges~\cite{Ran07}),
or the thermodynamic reasons discussed in this paper. Most probably at least 
two of these factors act simultaneously.

Another interesting result follows from Fig.~\ref{DGcurves}. The numbers
$n^*$ of atoms in the critical nuclei always satisfy either the relation
$n^* = i \times i + 1$ or, more frequently, $n^* = i \times (i - 1) + 1$,
where $i$ is an integer giving the number of atoms in the longer edge 
of the rectangle. This means that the critical nuclei consist of a 
compact structure plus one additional atom. In the
atomistic theory of nucleation on surfaces it is usually assumed that the
critical nuclei contain one atom less than those necessary for a 
compact structure~\cite{Wal69} (for a review see Ref. \onlinecite{Mar10}). 
Thus the smallest nucleus (larger than 1~atom) on a surface
with a square symmetry as in our case consists of 3 atoms located at the
vertices of a rectangular triangle. The smallest stable cluster consists 
of 4 atoms located at the corners of a square. In order to disrupt
a critical nucleus, one has to break a single bond whereas in order to 
disrupt the stable cluster, one has to simultaneously break two bonds. 
This explains the stability of the smallest stable cluster compared 
with the critical nucleus.

As noted by Kashchiev,~\cite{Kash08} the main difference between the classical
nucleation theory (CNT) and the atomistic nucleation theory (ANT) is the 
nucleus shape. 
Whereas ANT allows any irregular shape that arises from the atomistic
nature of the nucleus edges, in the CNT it is assumed that the nucleus
possesses the equilibrium shape, which in our case [(100) surface and 
consideration of only first-neighbors interactions] is a square island. 
What we observe is a rectangle (with the shape closest to a square) 
plus one additional atom. The compact shape can be understood in terms of
the highest coordination achieved. The question that
arises is about the additional atom which is the first in the new row.

Kaischew and Stranski derived ex\-pressions for the rate of 3D and 2D 
nucleation by using a completely kinetic approach.~\cite{Kai341,Kai342} 
They argued that the barrier of 
formation of a 3D crystalline nucleus should include the energy of 
formation of a 2D nucleus on the side wall of a 3D cluster smaller by
an atomic plane than the critical nucleus. Analogously, the barrier for 2D
nucleation should include the barrier of formation of a new atomic row, which
is in fact the barrier for attachment of the first atom of the row. The
single atom gives birth to a new row of atoms thus playing the role of
one-dimensional nucleus\cite{Vor70} (for a recent review see
Ref.~[\onlinecite{Mar03}]).

Considering the classical nucleation, Kaischew and Stranski also argued that
the work of formation of a 3D nucleus with a cubic shape should include the
work of formation of three 2D nuclei on neighboring crystal walls in order to
transform the cube of $i$ atoms in its edge to a cube of $i+1$ atoms thus 
preserving the equilibrium shape. Analogously, the 2D square
nucleus should build two rows of atoms on neighboring edges.

Kashchiev found that the nucleus size as a function of the supersaturation
always satisfies the relation $n^*~=~i~\times~(i-1)~+~1, (i = 2, 3, 4...)$,
which means that the single additional atom begins the atomic row that
transforms the rectangle $i\times(i-1)$ into an island with the square
equilibrium shape $i \times i$.~\cite{Kash08} The calculation in which the
work of formation of a 2D nucleus is corrected by including the contribution 
of the additional atom is in much better agreement with the exact solution 
for the nucleation rate as derived by Becker and D\"oring in their seminal 
paper.~\cite{Bec36} Kashchiev found that only rec\-tan\-gul\-ar 2D critical 
nuclei form; the reason is that he considered them to grow on an infinitely 
large surface from an infinite vapor phase. On the contrary, we simulate 
the construction of 2D nuclei on a strained layer of finite size 
by removing the atoms from the edges of the lower island. In such a case, 
the detailed atomistics at the lower island also play a role. 

\begin{figure}[ht]
\includegraphics*[width=8.5cm]{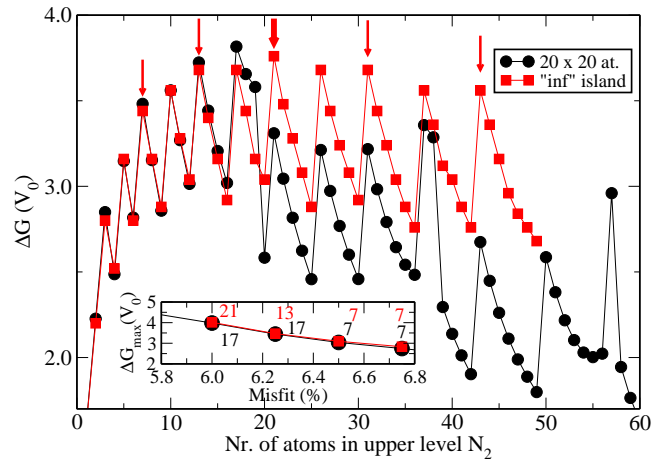}
\caption{\label{infinite}
Close up of the transformation curve for a lattice misfit of +3.5\%,
$\mu = 16$, $\nu = 14$ and a finite cluster of a size of 400 atoms
(circles). Also shown (squares) is the transformation curve for a
similar situation for which an infinitely large cluster was simulated
by not removing atoms from the first level but by correcting instead
for the binding energy at the half-crystal position at the center
of an atomic row (see text for details). The thick arrow marks the absolute
maximum of the transformation curve. The remaining arrows show that local
maxima of the latter curve tend to be higher for $i\times(i-1) + 1$ atoms,
with $i$ an integer, than for $i\times i + 1$ atoms.
The insert shows a selected region of the plot of the barrier height vs.
lattice misfit for a potential with $\mu = 2 \nu = 12$ containing also
the critical nucleus sizes for both types of configurations. These data
confirm the appearance of only ''rectangle + 1"-type islands for the
simulated infinite islands.}
\end{figure}

This is demonstrated by the results shown in Fig.~\ref{infinite}. 
Here, total energy curves were cal\-cu\-lat\-ed both for a fi\-n\-ite-sized
initial island of 20$\times$20 atoms and for a situation that simulates the
same transformation sequence for an infinitely large initial 
monolayer island: The same configurations for the growing cluster in the 
second atomic level were considered, but in this second case, no atoms were 
detached from the first atomic level. The increasing total number of atoms 
was corrected by subtracting the calculated negative binding energy of an 
atom at a kink (half-crystal) position at the middle of the last atomic 
row of the initial monolayer island (in the limit of a very large island, 
this quantity gives the chemical potential of the overlayer material 
on the considered substrate).~\cite{Kos27,Stra28} 

A close look at the transformation curve for the simulated infinite island in 
Fig.~\ref{infinite} reveals that the maxima corresponding to 
``rectangle + 1" configurations, marked by arrows in the figure, are 
slightly but consistently higher than those for ``square + 1" configurations 
when considered above the overall smooth curve described by the local
maxima. In this way, only the former configurations will evolve into global
maxima with varying lattice misfit.
This is ultimately a consequence of the fact that nuclei of rectangular shape 
have higher energies than those of square shape due to the square symmetry 
of the model geometry. 
This effect is more important if the absolute value of the misfit is large 
or the lower island is very big. Then the size $n^*$ of the second layer 
nucleus is much smaller than the number of atoms in the lower island edge 
and the atomistics of the latter does not play any significant role. 
This is further confirmed by the insert of Fig.~\ref{infinite}, which shows
the values at the maxima of the transformation curves, $\Delta G_{max}$ as
a function of the lattice misfit, together with the sizes of the critical
nuclei. It can be seen that the values of the barrier heights are very similar
for the finite-sized and for the simulated infinite islands, while
the critical sizes of the form ``square + 1" [($i\times i + 1$ atoms)] 
change to ``rectangle + 1" [$i\times(i-1) + 1$ atoms].

In summary, we have found that the work of formation of second layer nuclei 
in heteroepitaxy is smallest at the corner of the first-layer island and
largest at the center, in accordance with experiments. 
Thus thermodynamics of second layer nucleation is com\-pat\-ible with the
kinetics of this process as suggested by Villain.~\cite{Vil05} 
The critical nucleus consists of one atom in addition of a compact shape; 
It plays the role of a ``one-dimensional nucleus" giving rise to a new 
atomic row. The compact shape is in general a rectangle with edges of 
$i$ and $i-1$ atoms, while square shapes can also appear if the length of 
the critical nucleus is comparable to the number of atoms in the edge of 
the original first-layer island.

\acknowledgments This work was supported by the Spanish MiCInn, Project No. 
FIS2008-01431.

\end{document}